# Busting the Myth of Spontaneous Formation of $H_2O_2$ at the Air–Water Interface: Contributions of the Liquid–Solid Interface and Dissolved Oxygen Exposed

Muzzamil Ahmad Eatoo & Himanshu Mishra*

Environmental Science and Engineering (EnSE) Program,

Biological and Environmental Science and Engineering (BESE) Division,

King Abdullah University of Science and Technology (KAUST),

Thuwal, 23955-6900, Kingdom of Saudi Arabia,

Water Desalination and Reuse Center (WDRC),

King Abdullah University of Science and Technology (KAUST), Thuwal, 23955-6900, Kingdom of Saudi Arabia

Center for Desert Agriculture (CDA),

King Abdullah University of Science and Technology (KAUST), Thuwal, 23955-6900, Kingdom of Saudi Arabia

*Himanshu.Mishra@kaust.edu.sa

**Abstract:**

Recent reports on the spontaneous formation of $H_2O_2$(aq) at the air–water interface and the solid–water interface have been sensational. The speculated mechanism at the air–water interface is based on instantaneous ultrahigh electric fields and the 'micro' scale of droplets, whereas the solid–water interface is speculated to be the site for oxidation of water (or hydroxide ions) and reduction of the solid surface. We utilized $^1$H-NMR spectroscopy to investigate the effects of the nebulizing gas, the dissolved oxygen content, and solid substrates on the $H_2O_2$(aq) formation (detection limit ≥50 nM). Experiments revealed that, contrary to the sensational claims, the air–water interface is not the site for $H_2O_2$(aq) formation; instead, it's the solid–water interface where $H_2O_2$(aq) is formed during the reduction of dissolved oxygen and oxidation of the solid surface. Curiously, the tendencies of solid substrates towards forming $H_2O_2$(aq) follow the classic Galvanic series. This report advances the current understanding of aquatic chemistry and should be relevant to corrosion science, surface science, and electrochemistry.

**Introduction**

Hydrogen peroxide ($H_2O_2$) is an industrial chemical serving a wide range of applications, such as disinfection[1], chemical synthesis[2], rocket propulsion[3], and wastewater treatment[4]. Current production of $H_2O_2$ at scale relies on the anthraquinone cycling process that requires significant energy and generates waste[5], necessitating sustainable alternatives. Recent reports on the laboratory-scale production of $H_2O_2$ via electrochemical oxygen reduction are promising[6, 7, 8]. However, electrochemical methods are not devoid of shortcomings – the process is complex and ridden with the danger of spontaneous combustion and explosion. Multiple side reactions may further limit the scalability of the process. In this context, sensational reports on the spontaneous formation of $H_2O_2$ at the aerial interface of water microdroplets seem tantalizing[9, 10, 11, 12, 13]. Specifically, ~30 µM $H_2O_2$ was found in water microdroplets of diameter ≤ 20 µm sprayed via pressurized gas[9]; and ≤ 115 µM $H_2O_2$ was found in condensed water microdroplets on common substrates in the relative humidity range 40–70%[10]. The presence of ultrahigh electric fields on microdroplets' surface has been speculated to be the underlying cause[9, 10] and implications for atmospheric chemistry[14], human health and bactericidal applications[13], green chemistry[9, 14], and seasonality of diseases due to the Goldilocks effect[12, 15] have been pointed out. We introduce our investigation of this chemical transformation by noting that the interrogation of water's interfaces is notorious for artifacts arising from contamination, incorrect interpretations of experiments, and the lack of encompassing multiscale computational models[16, 17, 18, 19, 20, 21, 22, 23, 24, 25, 26, 27, 28, 29, 30, 31, 32, 33, 34, 35, 36, 37, 38]. Our experience in microdroplet chemistry has taught us to curb our enthusiasm until after the conclusion has been stress-tested via multiple experimental techniques[39].

The prospect of aerosolized water microdroplets producing $H_2O_2$ is appealing due to its greenness and potential ease of application[13, 40]. While some experiments[41, 42, 43, 44], computer simulations[45, 46], and a *Gedankenexperiment*[47] have given credence to the contention of spontaneous $H_2O_2$ formation at the air–water interface, others have disagreed[48, 49, 50, 51, 52, 53, 54, 55]. We commenced our investigation in 2019 by comparing the various commercially available assays for $H_2O_2(aq)$. Compared to the potassium titanium oxalate assay (PTO; detection limit ≥10 µM), utilized in the original reports[9, 10], the Hydrogen Peroxide Assay Kit (HPAK) affords a 40-times lower detection limit (≥250 nM). Equipped with HPAK, we utilized a glove box to interrogate $H_2O_2(aq)$ concentrations in condensed water microdroplets generated in $N_2(g)$ environment by gently heating water (50−70°C). This experiment revealed that the $H_2O_2$ concentrations in the bulk water and the condensates were indistinguishable[48]. We also found that if the condensates were produced via ultrasonic humidifiers, ~1 μM $H_2O_2$(aq) was produced – in the water reservoir, the mist, and the condensates[48]. This was due to cavitating bubbles formed under ultrasonic acoustic pressure in bulk, which produces OH• radicals[50, 56, 57, 58, 59]. However, we remained puzzled why Zare & co-workers found ~110 μM $H_2O_2$ in their experiments.

During 2020–2021, we broadened our investigation to include microdroplets produced by pneumatic sprays. This device – like the one in the original report[9] – facilitated the gas flow speeds of 100–1000 ms$^{-1}$, breaking up water droplets to form sprays. In all these studies, (i) sprayed water microdroplets were collected in glass bottles, and (ii) condensed microdroplets were formed on $SiO_2$/Si wafers. We discovered that ppm level of spontaneous $H_2O_2$ (1 ppm = 29.4 μM $H_2O_2$) formation took place only in the presence of $O_3$(g)[49].

In their latest report, Zare & co-workers[60] repeated spray experiments in a controlled gas environment and utilized NMR to quantify $H_2O_2$(aq) at 40 nM resolution following the protocol of Bax & co-workers (Bruker 600 MHz Avance III, non-cryogenic probe, 20,000 scans with 0.1 s acquisition time)[42, 61]. When water was injected through copper tubing in the flow range of 25–150 μL/min via a pressurized $N_2$ at 100 psi (6.8 atm), the $H_2O_2$(aq) concentration in the sprays ranged from 1.5–0.3 μM (95–99% reduction from the original report[9]). Interestingly, they also found that for a fixed liquid flow rate, as the nebulizing gas (fixed line pressure) was changed from (i) $N_2$ to (ii) $N_2$+$O_2$(2%) to (iii) $N_2$+$O_2$(21%) to (iv) $O_2$(100%), the $H_2O_2$(aq) concentration increased from (i) 0.49 ± 0.05 μM, (ii) 0.69 ± 0.05 μM, (iii) 1.12 ± 0.02 μM, and (iv) 2.00 ± 0.05 μM, respectively[60]. Based on these observations, they contend that their original claims were correct, i.e., microdroplets' air–water interface spontaneously produces $H_2O_2$. Here, even if we assume that the latest claim is valid, one has to admit that the previous reports[9, 10], which utilized the PTO assay ($H_2O_2$ detection limit ≥10 μM), could not have detected the 0.30–2.00 μM $H_2O_2$(aq) concentrations, i.e., they were reporting artifacts of the ambient ozone gas[49]. This admission may even help us understand why the latest[60] and previous[9] reports of Zare & co-workers present contradictory trends in the $H_2O_2$(aq) concentrations in water microdroplets when the concentration of dissolved oxygen ($O_2$(aq)) is increased. Lastly, two new experimental reports have surfaced wherein the $H_2O_2$ formation is also observed at the silica–water interface when (i) liquid water was passed through a PDMS microfluidic chip placed on glass[11], and (ii) water vapor was passed through a packed bed of $SiO_2$ nanoparticles[15]. As for the mechanism, the authors state, "*In fact, our proposed mechanism is built around the hypothesis that the overlap between the electron clouds of the* *water molecule and the solid surface during the contact will lead to the generation of $H_2O^+$ and $OH^-$*"[11]. They also present an example as follows, "*Then, the electron may transfer from the water molecule to the surface of SiO$_2$, which is the so-called contact electrification*"[15].

In this contribution, we investigate whether the skin of water – the air–water interface – is so unstable that airborne microdroplets can spontaneously produce $H_2O_2$, or whether something else is going on. We investigate the origins of ~1 µM $H_2O_2$ in condensed and sprayed water microdroplets via $^1$H-NMR, to answer the following interrelated fundamental questions:

1. Is it true that the $H_2O_2$ formation in water microdroplets is influenced by the nature of the nebulizing gas, viz., $N_2$ or $O_2$?
2. Would the $H_2O_2$ concentration in condensates collected in an inert gaseous environment be the same or different if the solid surface composition varied (e.g., SiO$_2$/Si wafer or stainless steel)?
3. What if we immersed pellets of a solid material (e.g., aluminum or mild steel) in bulk water or sandwiched a film of water between two solid surfaces to eliminate the air–water interface from the picture practically? In other words, is the 'micro' scale of droplets necessary for the spontaneous formation of $H_2O_2$ at aqueous interfaces?
4. What is the role of dissolved oxygen (in water) in this chemical transformation? If we removed the dissolved oxygen from water and sprayed it, would $H_2O_2$ still form?
5. During the spontaneous $H_2O_2$ formation at the solid–water interface, do water molecules transfer electrons to the solid and therefore reduce it?[11, 15]
6. Which aqueous interface would produce more $H_2O_2$ for a fixed area: the air–water interface or the solid–water interface? (Note: by solid we refer to common materials such as glass, steel, etc.)

**Results**

In a controlled gaseous environment ($N_2$(g), unless specified) afforded by a clean glovebox (Fig. S1), we collected water microdroplets formed via pneumatic sprays or by condensing the vapor generated by gently heating water (60 ℃) onto cold surfaces. First, we probed the effects of the nebulizing gas ($N_2$ or $O_2$) on the $H_2O_2$(aq) concentration. We utilized the flow rates suggested by Zare & co-workers[60] that maximized the $H_2O_2$ formation: water flow rate of 25 µL/min through a 0.10 mm-wide silica capillary, nebulizing $N_2$(g) gas at 100 psi shearing through an outer concentric tube of 0.43 mm diameter (Fig. 1a & S1). The spray was collected in a custom-built glass container described previously[49] (Fig. S1). The quantification of $H_2O_2$(aq) relied on $^1$H-NMR via the remarkable protocol developed by Bax & co-workers[61], which was also followed by Zare & co-workers[42, 60]. Briefly, we utilized a Bruker 950 MHz Avance Neo NMR spectrometer equipped with a 5mm Z-axis gradient TCI cryoprobe at 275K. During each measurement, a 6 ms Gaussian 90-degree pulse was applied to selectively excite the protons of $H_2O_2$, followed by a 53 ms acquisition time corresponding to 1024 detection points with a spectral width of 9615 Hz. Over 50,000 scans were collected with a recycle delay of 1 ms between the scans. With this technique, we could observe $H_2O_2$(aq) down to ~50 nM detection limit (See Fig. S2 for a representative calibration plot). The $^1$H-NMR results confirmed the presence of 1.0±0.2 μM $H_2O_2$ in water microdroplets sprayed using $N_2$ gas (Fig. 1). Next, as we switched the nebulizing gas to $O_2$, keeping the water flow the same, the $H_2O_2$ concentration increased to 3.0±0.2 μM (Fig. 1). Here, two crucial questions arise, which we next address: (i) If this phenomenon is driven by ultrahigh instantaneous electric field at the air–water interface, then why does the nebulizing gas influence it; (ii) Could the solid–water interface drive this chemical transformation?

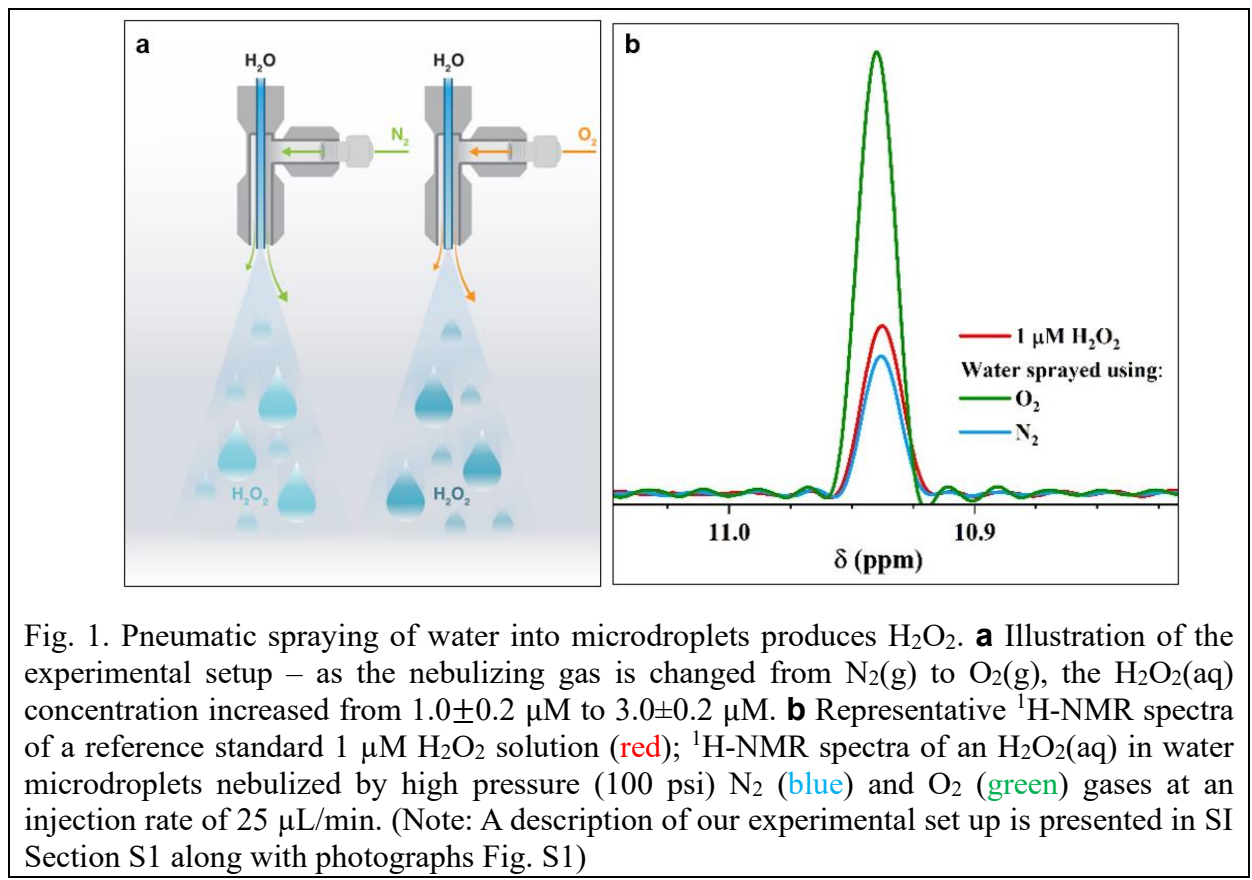

Fig. 1. Pneumatic spraying of water into microdroplets produces $H_2O_2$. **a** Illustration of the experimental setup – as the nebulizing gas is changed from $N_2$(g) to $O_2$(g), the $H_2O_2$(aq) concentration increased from 1.0±0.2 μM to 3.0±0.2 μM. **b** Representative $^1$H-NMR spectra of a reference standard 1 μM $H_2O_2$ solution (red); $^1$H-NMR spectra of an $H_2O_2$(aq) in water microdroplets nebulized by high pressure (100 psi) $N_2$ (blue) and $O_2$ (green) gases at an injection rate of 25 μL/min. (Note: A description of our experimental set up is presented in SI Section S1 along with photographs Fig. S1)

To answer those questions, we compared the amount of $H_2O_2$(aq) formed in water microdroplets condensed onto a variety of smooth and flat substrates, namely $SiO_2$/Si wafer, polished titanium, polished stainless steel (SS304), polished mild steel, silicon surface (obtained by reactive ion etching of SiO$_2$/Si wafers), polished copper (Cu), polished magnesium alloy (AZ31B), and polished aluminum (Al) (see Methods for details). Note: mechanical polishing was done with emery papers of grit size ranging from 400 to 1500 to remove the native oxide layer. While the size distribution of the microdroplets formed on those substrates did not vary significantly because of their superhydrophilic nature, there was a dramatic difference in the amount of H$_2$O$_2$(aq) in the condensates depending on the nature of the substrate(Fig. 2). For instance, as we replaced SiO$_2$/Si wafer substrate by Mg alloy (AZ31B), the H$_2$O$_2$ concentration shot up from 0.4±0.2 µM to 68±5 µM (Fig. 2). In our four years-long investigations of this phenomenon, replete predominantly with null results, this was the first time we observed the formation of ppm-level H$_2$O$_2$ (aq) in water microdroplets in the absence of O$_3$(g).

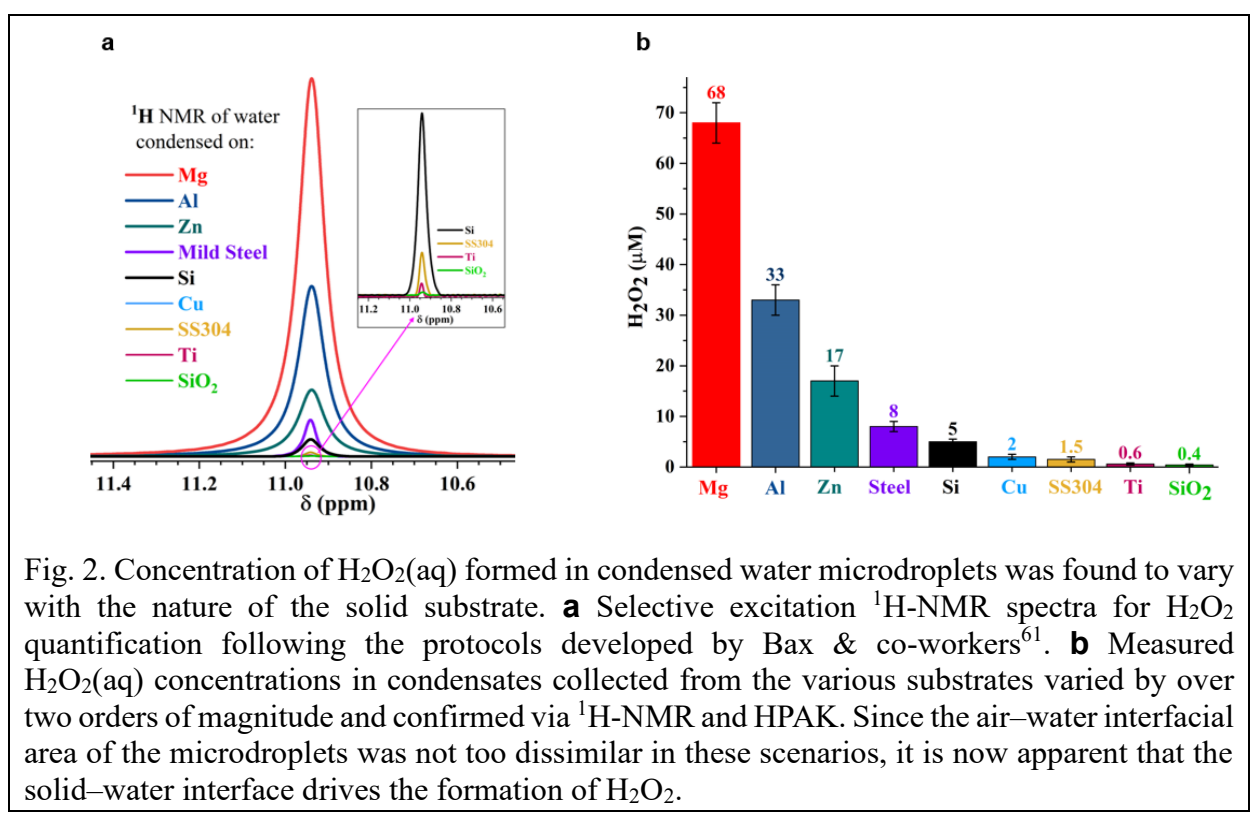

Fig. 2. Concentration of H$_2$O$_2$(aq) formed in condensed water microdroplets was found to vary with the nature of the solid substrate. **a** Selective excitation $^1$H-NMR spectra for H$_2$O$_2$ quantification following the protocols developed by Bax & co-workers[61]. **b** Measured H$_2$O$_2$(aq) concentrations in condensates collected from the various substrates varied by over two orders of magnitude and confirmed via $^1$H-NMR and HPAK. Since the air–water interfacial area of the microdroplets was not too dissimilar in these scenarios, it is now apparent that the solid–water interface drives the formation of H$_2$O$_2$.

Notably, the H$_2$O$_2$ concentrations reported in Fig. 2 were those obtained on freshly prepared surfaces, i.e., without a native oxide layer; over time, as the condensation experiments were repeated on the same surface, the extent of H$_2$O$_2$(aq) formation decreased, underscoring the importance of the solid–water interface. For instance, the H$_2$O$_2$ produced on freshly prepared Al surface was around 30–35 µM, which decreased to 7–8 µM in the second cycle and 4–5 µM in the 3$^{rd}$ cycle (each separated by 10 mins).

Having identified that microdroplets placed on common materials like (polished) aluminum produce ppm-level $H_2O_2$, we probed the importance of the droplet size in this chemical transformation. We formed a 1:1 volumetric solution of DI water with the HPAK reaction mixture and placed a macroscopic 1000 μL droplet (of base diameter of 12000 μm) onto an aluminum plate. Within a few seconds, we observed a sharp blue fluorescence – proof for the formation of $H_2O_2$(aq)– with an unambiguous gradient emanating from the Al–water interface (Fig. 5 and SI Movie 1). Judging by the fluorescence intensity, the local concentration of $H_2O_2$(aq) at the solid–liquid interface is at ppm-level, i.e., while the air–water interface produced no fluorescence visible to the naked eye.

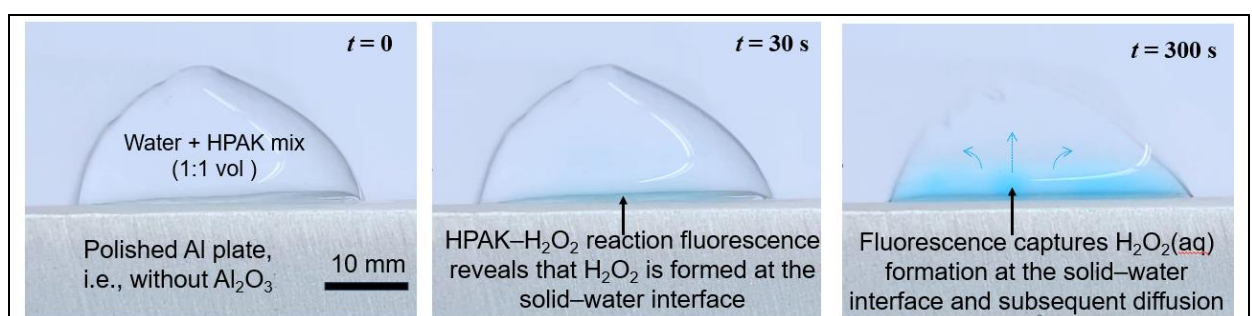

Fig 3. Time-dependent formation of $H_2O_2$(aq) in a macroscopic droplet of 1:1 mixture of water and HPAK reaction mixture placed onto an aluminum plate (see SI Movie S1). Within seconds, $H_2O_2$(aq) formation at the aluminum–water interface is apparent. This simple experiment proves that the size of the droplet and the air–water interface do not matter; in fact, it's the solid–water interface that drives this chemical transformation. Note: Al surface is superhydrophilic, and water spreads on it as a film; so, we placed Al plate vertically on a polystyrene sheet and formed a 1 ml droplet resting onto the Al edge. Also, the 1:1 mixture of water and HPAK reaction on polystyrene did not yield the faintest blue fluorescence visible to the naked eye.

Building on this experiment, we reduced the air-water interfacial area from this three–phase system by (i) sandwiching fresh 1ml of DI-water–HPAK 1:1 mixture between two 20×20 cm$^2$ aluminum plates; and (ii) immersing freshly polished Al (or Mg) pellets to 5ml bulk DI-water–HPAK 1:1 mixture (Fig. 4c–d). In all these scenarios, we discovered that the solid–water interface was the site for the spontaneous $H_2O_2$ formation, and the air–water interface had negligible effect, if any. This observation is based on the distinct color gradient at the solid–water interface. After these experiments, we collected the water samples (films or bulk) and did $^1$H-NMR measurements (Table S1). A systematic study of the effects of the solid–water surface area and the effect of time is underway. These results unambiguously establish that the spontaneous $H_2O_2$ production in water does not necessitate microscopic droplets or the air–water interface, and it can take place even in bulk water when specific solid materials are introduced.

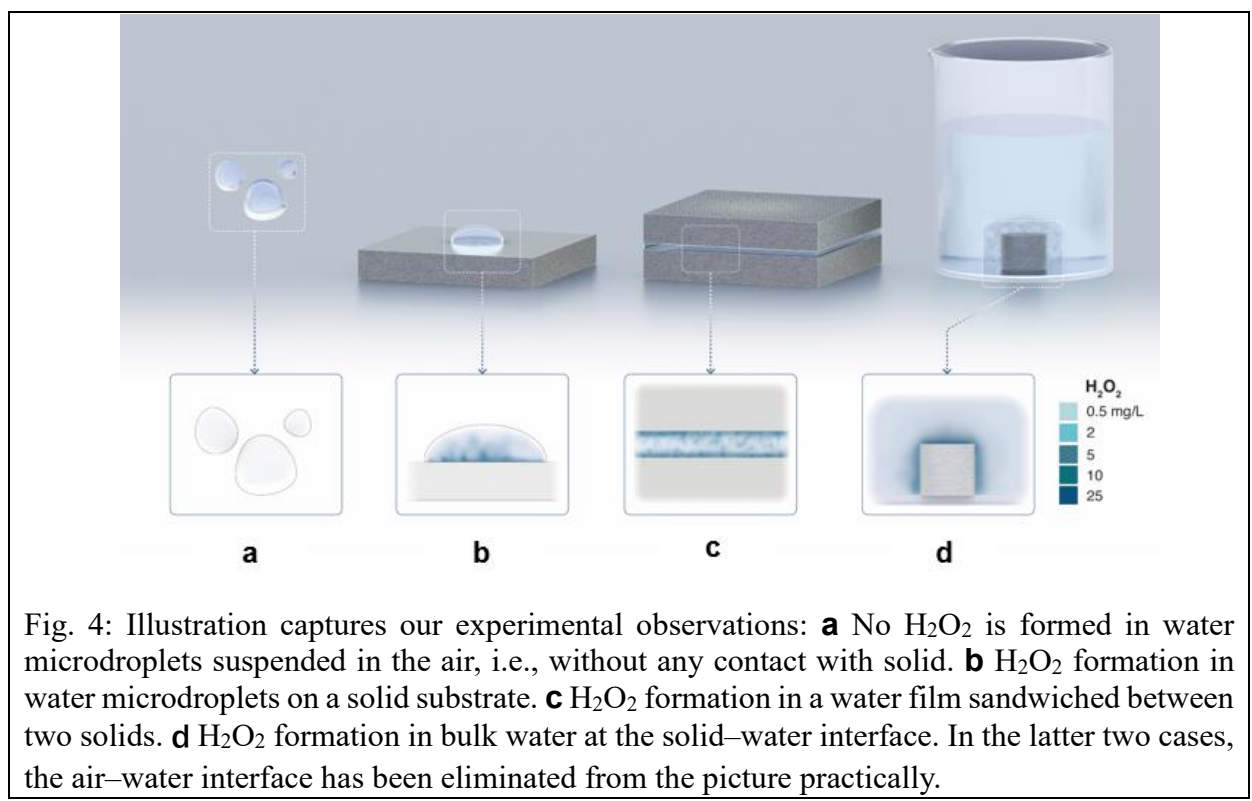

Fig. 4: Illustration captures our experimental observations: **a** No $H_2O_2$ is formed in water microdroplets suspended in the air, i.e., without any contact with solid. **b** $H_2O_2$ formation in water microdroplets on a solid substrate. **c** $H_2O_2$ formation in a water film sandwiched between two solids. **d** $H_2O_2$ formation in bulk water at the solid–water interface. In the latter two cases, the air–water interface has been eliminated from the picture practically.

Next, we probed the surfaces before and after contact with water via X-ray photoelectron spectroscopy (XPS) (Figs. 2–4). This revealed that the spontaneous $H_2O_2$ formation at the water-solid surface was accompanied by the oxidation of the substrate (Fig. 3). For instance, on contact with water (condensed or both), metallic aluminum ($Al^0$) was oxidized to $Al^{3+}$, and semiconductor silicon ($Si^0$) was oxidized to $Si^{4+}$ oxidation state (Fig. 5a–c). Note: water microdroplets spread and merged during each cycle, covering the entire (superhydrophilic) solid surface by the end of each cycle. This finding contradicts the claims that during $H_2O_2$ formation at the solid–water interface, $OH^-$ ions get oxidized to $OH^·$ (or $H_2O$ molecules get oxidized to $H_2O^+$)[11, 15], because if this were true, the solid surface would be getting reduced, which is not the case. Note: oxidation product characterization is beyond the scope of this study and will be reported in the future.

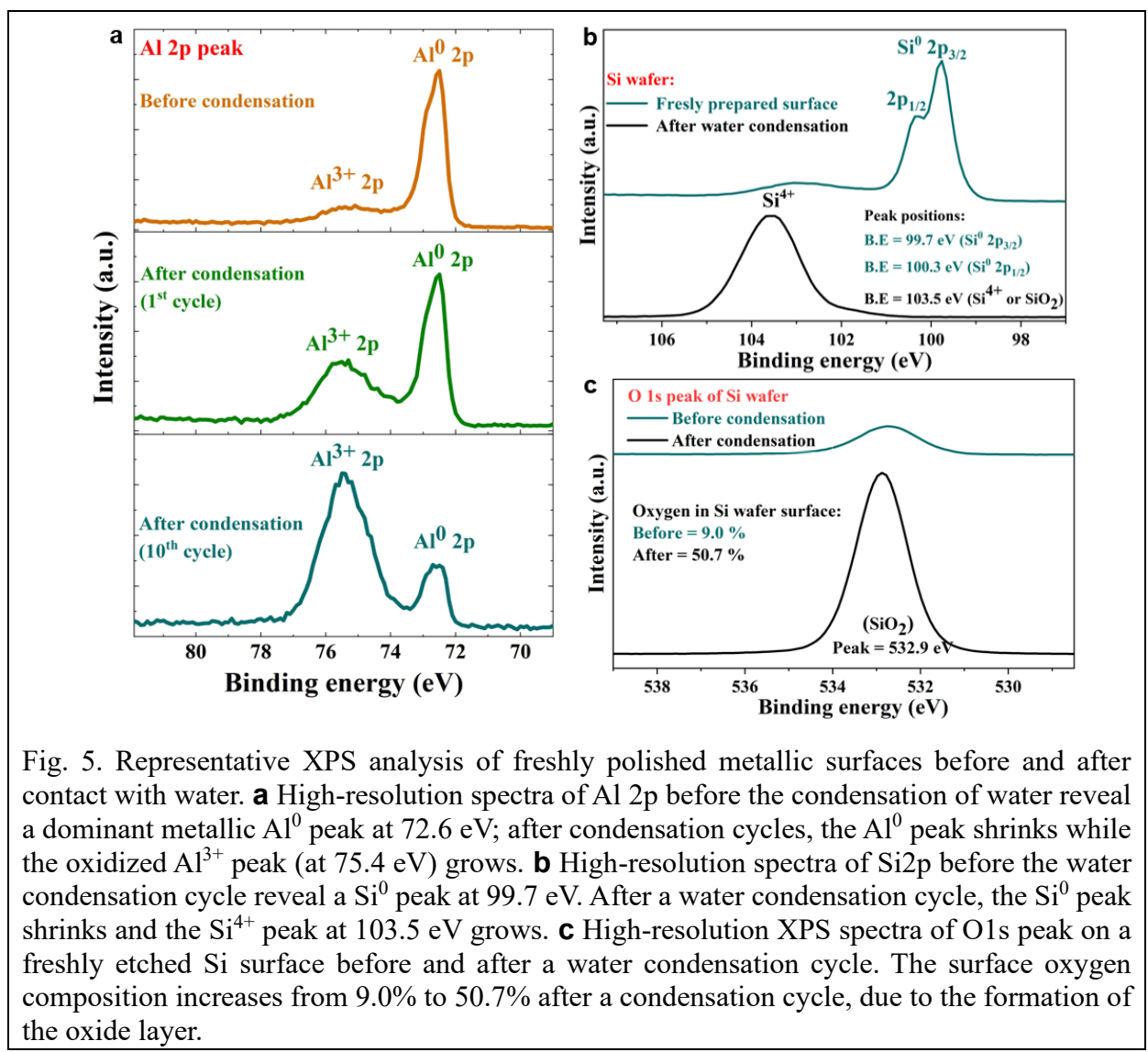

Fig. 5. Representative XPS analysis of freshly polished metallic surfaces before and after contact with water. **a** High-resolution spectra of Al 2p before the condensation of water reveal a dominant metallic $Al^0$ peak at 72.6 eV; after condensation cycles, the $Al^0$ peak shrinks while the oxidized $Al^{3+}$ peak (at 75.4 eV) grows. **b** High-resolution spectra of Si2p before the water condensation cycle reveal a $Si^0$ peak at 99.7 eV. After a water condensation cycle, the $Si^0$ peak shrinks and the $Si^{4+}$ peak at 103.5 eV grows. **c** High-resolution XPS spectra of O1s peak on a freshly etched Si surface before and after a water condensation cycle. The surface oxygen composition increases from 9.0% to 50.7% after a condensation cycle, due to the formation of the oxide layer.

Following the XPS study, we noticed that the $H_2O_2$ formation was accompanied by surface oxidation. Therefore, we got curious whether it was due to the reduction of the dissolved oxygen.[62] To examine this, we first removed dissolved oxygen from water by heating it in an autoclave until its boiling point, followed by $N_2(g)$ bubbling for 45 minutes and then sealing it inside an $N_2$-purged (Methods). This treatment reduced the $O_2(aq)$ concentration to < 0.01 mg/L. Now, microdroplets of oxygen-free water were formed via pneumatic spraying using $N_2(g)$ in an $N_2$ environment and collected in glass containers (following the same protocol as before), and the $H_2O_2(aq)$ concentration was compared with that in the microdroplets formed with water containing dissolved $O_2(g)$ (Fig. 6a). Remarkably, in the absence of $O_2(aq)$, we did not observe any $H_2O_2(aq)$ via 1H-NMR (detection limit ≥ 50 nM) (Fig. 6B). Next, we tested the effects of dissolved $O_2(g)$ on the formation of $H_2O_2(aq)$ in bulk water by adding metallic pellets (Mg or Al). Here, we tested the following three scenarios:

(i) A Mg pellet was added to water saturated with the ambient $O_2(g)$, and the vial was left open in an ozone-free ambient environment.

(ii) A Mg pellet was added to water saturated with ambient oxygen $O_2(g)$ and then sealed.

(iii) A Mg pellet was added to oxygen-free water in an $N_2(g)$ environment and sealed.

Curiously, the vial s open to the ambient air had significantly higher $H_2O_2(aq)$ than that in the sealed vial containing water saturated with dissolved $O_2(g)$. This means that the formation of $H_2O_2$ at the solid–water interface consumes dissolved $O_2(g)$, i.e., it's the limiting factor. We also characterized the consumption of the dissolved $O_2(g)$ before and after adding the pellets and found it to decrease over time (Fig. S4). Notably, in the absence of dissolved $O_2(g)$, we did not observe $H_2O_2(aq)$ within the detection limit of 50 nM (Fig. 6b). These results unambiguously establish (i) the importance of dissolved $O_2(g)$ in this chemical transformation; and (ii) the air–water interface of microdroplets is incapable for forming $H_2O_2$.

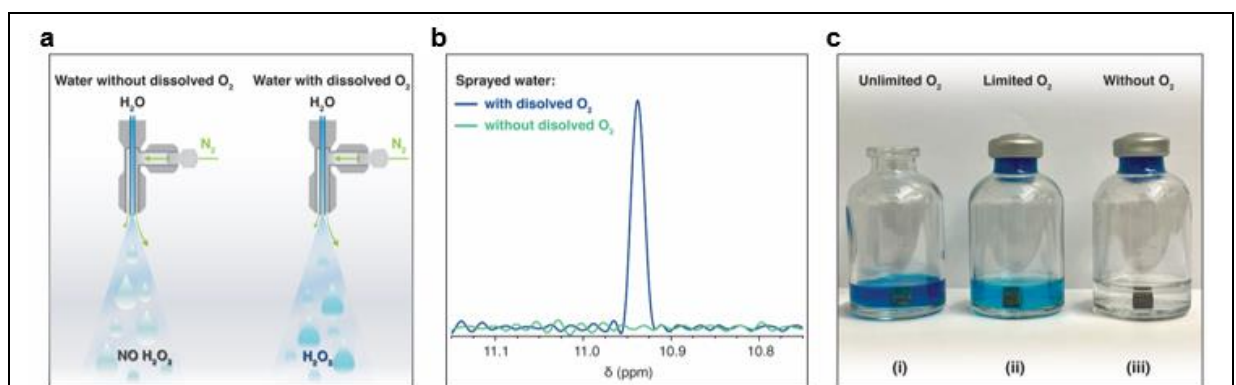

Fig. 6: Role of the dissolved oxygen in water on the formation of $H_2O_2$ in microdroplets and bulk forms. **a** An illustration of the experiments wherein water containing dissolved oxygen and deoxygenated were sprayed to form microdroplets. The microdroplets were then collected in a glass vial, and $H_2O_2(aq)$ was quantified. **b** Within a detection limit of 50 nM, $^1$H-NMR revealed that no $H_2O_2$ formed in the deoxygenated water, whereas $H_2O_2(aq)$ was readily detected in the presence of dissolved oxygen. Since the air–water interface was common in both scenarios, these experiments prove that the $H_2O_2(aq)$ formation happens at the solid–water interface due to the reduction of dissolved $O_2(aq)$ and the oxidation of the surface. Note: the nebulizing gas enhances the concentration of $H_2O_2(aq)$ by evaporating water during the collection process. **c** In another experiment, we prepared a 1:1 mixture of the HPAK reaction mixture with deoxygenated water and used a 1:1 mixture of HPAK reaction mixture with water containing dissolved $O_2(aq)$ as control. Next, $H_2O_2$ formation in the following three scenarios was investigated: (i) an Mg pellet was added to the 1:1 mixture saturated with the ambient $O_2(g)$ and the vial was exposed to the ambient air, i.e., 'unlimited oxygen case'; (ii) an Mg pellet was added to 1:1 mixture saturated with dissolved oxygen $O_2(g)$, and the vial was sealed, i.e., 'limited oxygen case'; and (iii) a pellet was added to the 1:1 mixture without dissolved $O_2(g)$, and the vial was sealed, i.e., 'without $O_2$ case'. Using HPAK, we found that in the absence of dissolved $O_2(aq)$, there was no $H_2O_2(aq)$ formed within the detection limit of 0.25

µM, whereas it appeared readily in the presence of $O_2$(aq). This demonstrates that this chemical transformation takes place at the solid–water interface, and dissolved $O_2$(aq) is a reactant. Therefore, $H_2O_2$ formation is not a property of the air–water interface or dependent on the size of the droplets. (Scale bar: the diameter of the pellet is 1 cm)

**Discussion**

Here, we draw together the results of this study and previous scientific reports (Fig. 7) and discuss the mechanisms underlying the formation of $H_2O_2$ in interfacial water. Our first finding is that the amount of $H_2O_2$(aq) formed in water condensates (or sprayed microdroplets) depends only on the nature of the surface on which it is collected (Figs. 2–3). In other words, the air–water interface or the size of microdroplets has no bearing on the $H_2O_2$(aq) formation (Figs. 3, 4d, and S5). For instance, as the air–water interface is reduced/eliminated from the picture, via sandwiching water films between solid plates (Fig. 4c) or by introducing solid pellets into bulk water, the formation of $H_2O_2$(aq) remains unaffected (Figs. 4d and S5). Our second crucial finding is that if the dissolved oxygen is removed from the water, there is no evidence for $H_2O_2$(aq) formation within our detection limit (Fig.6a–c). This observation contradicts the speculative mechanism for $H_2O_2$(aq) formation due to charge transfer between positively ($H_3O^+$ rich) and negatively charged ($OH^-$ rich) microdroplets[43, 47]. Next, our XPS results demonstrate that the formation of $H_2O_2$(aq) is accompanied by the oxidation of the solid surface and the reduction of dissolved oxygen (Fig. 5) (Fig. S4 shows how the absolute concentration of $O_2$(aq) decreases during $H_2O_2$ formation). When we examined the various commercially available materials utilized in this study in terms of their ability to form $H_2O_2$(aq) in (air-equilibrated) water, the trend followed the Galvanic series: Mg > Al > Zn > mild steel > Si > Cu > stainless steel (SS304) > Ti > $SiO_2$/Si wafer (Fig. 7). These findings therefore refute the previous speculations for the oxidation of $OH^-$ ions to $OH^•$ (or the oxidation of $H_2O$ to $H_2O^+$) and the reduction of the solid surface during the $H_2O_2$ formation at the solid–water interface[11, 15].

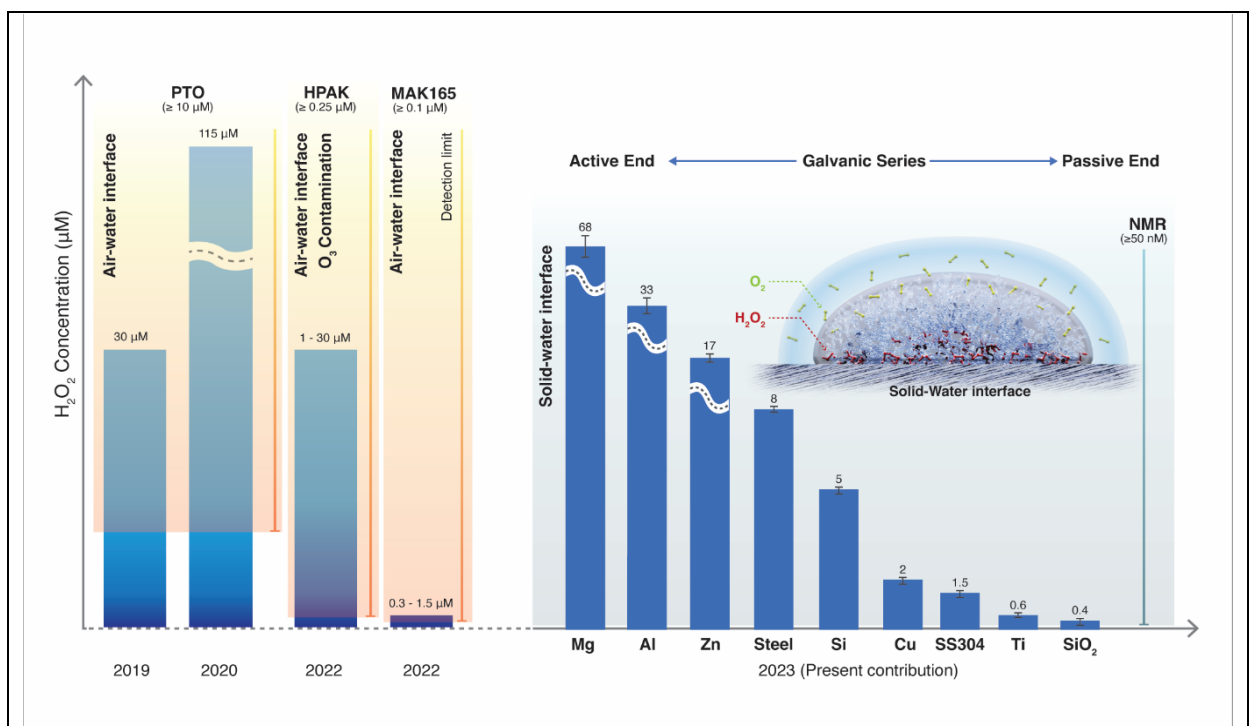

Fig. 7. Current understanding of the spontaneous $H_2O_2$ formation in water microdroplets since its first report in 2019. Zare & co-workers' initial reports used the PTO assay (detection limit ≥10 μM) and found 30 and 110 μM $H_2O_2$(aq) in sprays and condensates, respectively[9, 10]. In 2022, it was revealed using HPAK (detection limit ≥0.25 μM) that ambient $O_3$(g) can give rise to severe artifacts in these microdroplet experiments[49]. In 2023, using $^1$H-NMR (detection limit ≥0.04 μM), it was contended that in an ozone-free environment, the air–water interface still produces $H_2O_2$ (~ 1 μM)[60]. In the present contribution, we reveal using $^1$H-NMR that the solid–water interface is the site for $H_2O_2$(aq) formation, and the air–water interface does not contribute to $H_2O_2$ formation (quantified within the detection limit of ≥0.05 μM). Notably, if dissolved $O_2$(g) is removed from the water, $H_2O_2$(aq) is not observed within our detection limit. Next, we reveal that depending on the nature of the substrate that water contacts (as microdroplets or a film or as bulk water), the amount of $H_2O_2$(aq) formed follows the classic Galvanic series[63].

We postulate that the initiation of this chemistry involves the reduction of dissolved $O_2$(aq) by the solid surface, i.e., the surface transfers two electrons into interfacial $O_2$(aq), which transforms it into a highly reactive peroxide dianion ($O_2^{\bullet 2-}$)[64]. We anticipate this anion species to hover near the solid–water interface due to the electrostatic attraction. Next, the anion reacts with interfacial water molecules to form $H_2O_2$ and hydroxide ions[62, 64] (the reactions below capture this logic):

$$O_2 + 2e^- \rightarrow O_2^{\bullet 2-} \tag{1}$$

$$O_2^{\bullet 2-} + 2H_2O \rightarrow H_2O_2 + 2OH^- \tag{2}$$

Eq. (2) notes that the $H_2O_2$(aq) formation is accompanied by pH enhancement. We also observed that the pH of the condensate collected on the Mg plate was around 7.9±0.2, whereas the pH of the water reservoir used for gentle heating was 5.6±0.1. Understandably, the $H_2O_2$(aq) formation rate is the highest when the surface is free of native oxide and slows down as the oxide layer grows – also noticed in our experiments. Here, it is also important to note that similar reaction schemes have been proposed recently, and in some of them, $O_2$ is the byproduct[41]. If this were true, then the $H_2O_2$(aq) formation due to the addition of a metal pellet (Mg or Al) to bulk water would be the same whether dissolved oxygen content was (i) unlimited (Fig 6cii), (ii) limited (Fig 6cii) or (ii) nil (Fig 6cii), but that is not the case. Note: an in-depth investigation of the reaction intermediates (Eq. 1–2) and the contribution of metal pellets on water-splitting reactions and the water pH is underway.

**Conclusion:**

Our findings bust several myths surrounding the spontaneous formation of $H_2O_2$ at the air–water interface, including the instantaneous ultrahigh electric fields, the 'micro' scale of droplets[9, 10, 13, 41, 42, 45, 46, 60], and empirical arguments based on hydration enthalpies[47]. For water containing dissolved oxygen, which is commonplace in environmental and applied scenarios, the solid–water interface is the site where $O_2$(aq) reduces and forms $H_2O_2$(aq). Notably, the solid's ability to drive this chemical transformation depends on its position in the Galvanic series, e.g., Mg and Al have low oxidation resistance, so they form higher $H_2O_2$(aq). In contrast, Ti and stainless steel have high oxidation resistance, so they form lower $H_2O_2$(aq). Our XPS experiments reveal that the solid surface gets oxidized during the formation of $H_2O_2$. This refutes the speculation that during the formation of $H_2O_2$ at the solid–water interface, water molecules transfer electrons to the solid and reduce it[11, 15]. Crucially, in the absence of dissolved oxygen in water, $H_2O_2$(aq) was not observed in pneumatic sprays or in bulk water containing pellets of Mg or Al, down to the 50 nM detection limit. This proves that (i) the air–water interface of sprayed microdroplets and the putative (instantaneous) ultrahigh electric field therein are not capable of spontaneously forming $H_2O_2$ (Fig. 6); and (ii) the presence of dissolved oxygen is a required condition for the solid–water interface to form $H_2O_2$. We, therefore, submit that the latest claims[60] of the formation of 0.3–1.5 μM $H_2O_2$(aq) in water microdroplets (containing dissolved $O_2$) suffered from artifacts arising due to the unavoidable physical contact of water with solid surfaces, e.g., during sample preparation, collection, and analysis, as well as due to evaporative concentration[49]. Notably, when water microdroplets were formed by nebulizing with $O_2$(g), it increased $O_2$(aq) concentration, which promoted the formation of $H_2O_2$(aq) at the solid-water interface. Conversely, if the water is devoid of $O_2$(g), $H_2O_2$ is not formed spontaneously at the solid-water interface; here, the air–water interface can contribute to the $H_2O_2$ formation in the following two ways: (i) transfer $O_2$(g), which will be reduced at the solid–water interface to form $H_2O_2$(aq) (see inset in Fig. 7); or (ii) transfer $O_3$(g) that will oxidize water to form $H_2O_2$(aq) without the need of the solid–water interface. We hope these findings will advance the current knowledge of aquatic chemistry and prove to be relevant in corrosion science, electrochemistry, and surface science. Lastly, it has not escaped our attention that this chemistry and the subtle changes in the pH during the formation of $H_2O_2$(aq) may have a bearing on experiments that strive to interrogate whether protons or hydroxides have a higher propensity for aqueous interfaces.

**Materials and Methods:**

**Chemicals**

Deionized water obtained from a Milli-Q Advantage 10 set-up (18.2 MΩ-cm resistivity). Commercially available 30% hydrogen peroxide ($H_2O_2$) solution (Sigma-Aldrich CAS no. 7722-84-1) and deuterium oxide (($D_2O$, Catalogue no.3000007892) were used in this study.

**Spraying microdroplets**

Inside a glove box with a controlled $N_2(g)$ atmosphere to prevent ambient contamination, water was injected using a stainless steel capillary tube with inner diameter of 100 μm using a syringe pump (PHD Ultra, Harvard Apparatus). For nebulizing the water stream, ultra-pure nitrogen/oxygen was pushed through a coaxial stainless steel sheath with an inner diameter of 430 μm (Fig. S1). Liquid water flow rate was 25 μL/min and approximately 2 mL of microdroplets volume were collected in clean glass vials for further analysis.

**Deoxygenation of water**

Water was heated in an autoclave until its boiling point, followed by cooling via $N_2(g)$ bubbling for 45 minutes, which brought the temperature down to ~40℃. A dissolved oxygen sensor (WTW Multi 3320) measured the dissolved oxygen concentration in water with a detection limit of 0.01 mg/l. After this treatment, we could not see any signal for the O2(aq), meaning it was below our detection limit. Next, the water was quickly transferred into a glove box filled with N2 gas, where glass bottles were filled with it and sealed. The water sealed in an $N_2(g)$ filled glove box was autoclaved at 121 ℃ for 10 minutes to remove organic contamination on the vials.

**Substrates for condensation**

Silicon wafers of ~300 μm thickness, 4" diameter, and 2-μm-thick thermally grown oxide were purchased from Silicon Valley Microelectronics (Catalogue #SV010, p-type and 100 orientation), and we refer to them as $SiO_2$/Si wafer. Fresh Si surfaces were prepared by etching the $SiO_2$ layer via Reactive Ion Etching (using $C_4F_8$ and $O_2(g)$ for 5 minutes) inside the KAUST Cleanroom[65]. Right after this, condensation experiments were performed on the etched surfaces and water samples were collected in clean glass vials for $^1$H-NMR analysis. Next, the following commercially available plates comprised of metals and/or metallic alloys were utilized: Mg alloy (AZ31B, Thermo Scientific, Catalogue No. AA14066RF), Al plate (Fisher Scientific, Catalogue no. AA42124RF), mild-steel plate, stainless steel (SS304) plate, Zn plate (Thermo Scientific, Catalogue No. AA11914FI), Cu plate (Thermo Scientific Chemicals, Catalogue no. AA43822KS), and Ti (ASTM B 265 Trinity Brand Industries INC.™ part #6T-5). The native oxide on the metal plates was removed via mechanical polishing using silicon carbide emery papers of grit-size 400 to 1500 followed by cleaning with pressurized $N_2$ gas. For the condensation experiments, DI water (mentioned above) was heated at 60 ℃ inside a closed chamber to produce the water vapor. Water microdroplets formed onto cooled substrates (placed directly on ice) were collected using a low-pressure $N_2$ gas stream, and then transferred to NMR tubes and glass vials for further analysis.

**Hydrogen Peroxide Assay Kit (HPAK) assay**

$H_2O_2$ concentration inside the condensed water was quantified using the Hydrogen Peroxide Assay Kit (Fluorometric-Near Infrared, Catalogue # ab138886). It contains its unique AbIR Peroxidase Indicator that produces fluorescence independent of the solution pH in the range 4–10. Its maximum excitation wavelength is at 647 nm and maximum emission at 674 nm. Horseradish peroxidase enzyme catalyzes the reaction between $H_2O_2$ and the indicator and enhances the fluorescence signal. This facilitates the linear range of detection from 30 nM to 10 μM. The calibration curve (Fig. S3) was realized by adding 50 μL of an $H_2O_2$ standard solution from a concentration of 50 nM to 10 μM into 50 μL of the $H_2O_2$ reaction mixture using a black 96-well microtiter-plate, and the SpectraMax M3 microplate reader (Molecular Devices LLC). The analysis software used was SoftMax Pro 7. The water microdroplets were analyzed similarly by mixing 50 μL of each sample with the $H_2O_2$ reaction mixture, thus obtaining the respective concentration by the calibration curve.

**NMR Spectroscopy analysis and sample preparation**

No chemical was added to adjust the pH of samples to avoid contamination. In each case, 10 μL $D_2O$ was added to 490 μL analyte in a regular 5 mm quartz NMR tubes for the testing. All the NMR measurements were carried out on a Bruker 950 MHz Avance Neo NMR spectrometer equipped with a 5mm Z-axis gradient TCI cryoprobe at the temperature of 275K. During the measurement, a 6 ms Gaussian 90-degree pulse was applied to selectively excite the proton of Hydrogen Peroxide, followed by a 53 ms acquisition corresponding to 1024 detecting points with spectral width of 9615 Hz. Over 50,000 scans were collected with a recycle delay of 1 ms between scans. The NMR data were analysed by using TopSpin 4.2.0 software.

**XPS measurements**: Kratos Axis Supra instrument equipped with a monochromatic Al Kα X-ray source (hv = 1486.6 eV) operating at a power of 75 W and under UHV conditions in the range of ∼10−9 mbar was used to obtain the data. All the spectra were recorded in hybrid mode, using magnetic and electrostatic lenses and an aperture slot of 300 μm × 700 μm. The high-resolution spectra were acquired at fixed analyzer pass energies of 20 eV. The adventitious carbon (C 1s) peak at (284.5 eV) were used as a reference for calibration of all the peaks.

**Acknowledgments**

The co-authors thank Dr. Adair Gallo and Ms. Nayara Musskopf for building a robust experimental setup equipped with pneumatic sprays, a glove box, an ozone meter, etc., in HM's laboratory. The co-authors are indebted to KAUST's Dr. Xianrong Guo, Dr. Christian Canlas, Prof. Lukasz Jaremko, and Dr. Spyridon Gourdoupis for teaching them how to use 1H-NMR and giving them their own precious slots on the Bruker 950 MHz NMR spectrometer for over 3 months to support this research. The co-authors thank Mr. Heno Hwang, Scientific Illustrator at KAUST, for preparing illustrations in Figures 1, 4, 6, and 7; Mr. Amin Haider, Dr. Sankara Arunachalam, Dr. Hari Anand Rao, and Dr. Nimer Wehbe from KAUST for assistance with Fig. S1, the $SiO_2$ etching work, the water deaeration, and the XPS results respectively. The co-authors dedicate this paper to Prof. Rudy Marcus' centennial celebrations and thank him as well as Prof. Richard Saykally (UC Berkeley), Prof. Harry Gray (Caltech), Prof. Bill Goddard (Caltech), Prof. Paul Cremer (Penn State University), and Dr. Adair Gallo (Terraxy LLC) for fruitful discussions.

**Author contributions:** HM conceived the research plan and oversaw its execution. ME designed and performed the condensation and spray experiments inside the glovebox and collected the $^1$H-NMR and HPAK data. ME and HM analyzed the data and wrote the manuscript together.

**Competing interests:** The authors declare no competing interests.

**Data and materials availability:** All the data needed to evaluate the conclusions in the paper are present in the paper and/or the Supplementary Materials.

**Funding:** HM acknowledges KAUST for funding (Grant No. BAS/1/1070-01-01).

# References

1. Otter JA, Yezli S, Barbut F, Perl TM. 15 - An overview of automated room disinfection systems: When to use them and how to choose them. In: *Decontamination in Hospitals and Healthcare (Second Edition)* (ed Walker J). Woodhead Publishing (2020).

2. Kurti L, Czakó B. *Strategic Applications of Named Reactions in Organic Synthesis*. Academic Press; Illustrated edition (29 April 2005).

3. Kopacz W, Okninski A, Kasztankiewicz A, Nowakowski P, Rarata G, Maksimowski P. Hydrogen peroxide–A promising oxidizer for rocket propulsion and its application in solid rocket propellants. *FirePhysChem* **2**, 56-66 (2022).

4. Ksibi M. Chemical oxidation with hydrogen peroxide for domestic wastewater treatment. *Chemical Engineering Journal* **119**, 161-165 (2006).

5. Hans-Joachim R, Georg P. Production of hydrogen peroxide.). Google Patents (1939).

6. Murray AT, Voskian S, Schreier M, Hatton TA, Surendranath Y. Electrosynthesis of Hydrogen Peroxide by Phase-Transfer Catalysis. *Joule* **3**, 2942-2954 (2019).

7. Zhang X*, et al.* Electrochemical oxygen reduction to hydrogen peroxide at practical rates in strong acidic media. *Nature Communications* **13**, 2880 (2022).

8. Wang K, Huang J, Chen H, Wang Y, Song S. Recent advances in electrochemical 2e oxygen reduction reaction for on-site hydrogen peroxide production and beyond. *Chemical Communications* **56**, 12109-12121 (2020).

9. Lee JK*, et al.* Spontaneous generation of hydrogen peroxide from aqueous microdroplets. *Proceedings of the National Academy of Sciences of the United States of America* **116**, 19294-19298 (2019).

10. Lee JK*, et al.* Condensing water vapor to droplets generates hydrogen peroxide. *Proceedings of the National Academy of Sciences of the United States of America* **117**, 30934-30941 (2020).

11. Chen B*, et al.* Water–solid contact electrification causes hydrogen peroxide production from hydroxyl radical recombination in sprayed microdroplets. *Proceedings of the National Academy of Sciences* **119**, e2209056119 (2022).

12. Dulay MT, Huerta-Aguilar CA, Chamberlayne CF, Zare RN, Davidse A, Vukovic S. Effect of relative humidity on hydrogen peroxide production in water droplets. *QRB Discovery* **2**, e8 (2021).

13. Dulay MT, Lee JK, Mody AC, Narasimhan R, Monack DM, Zare RN. Spraying Small Water Droplets Acts as a Bacteriocide. *QRB Discovery* **1**, e3 (2020).

14. Zhu C, Francisco JS. Production of hydrogen peroxide enabled by microdroplets. *Proceedings of the National Academy of Sciences* **116**, 19222-19224 (2019).

15. Xia Y*, et al.* Contact between water vapor and silicate surface causes abiotic formation of reactive oxygen species in an anoxic atmosphere. *Proceedings of the National Academy of Sciences* **120**, e2302014120 (2023).

16. Mishra H*, et al.* Bronsted basicity of the air-water interface. *Proceedings of the National Academy of Sciences of the United States of America* **109**, 18679-18683 (2012).

17. Saykally RJ. Air/water interface: Two sides of the acid-base story. *Nature Chemistry* **5**, 82-84 (2013).

18. Gallo A*, et al.* The chemical reactions in electrosprays of water do not always correspond to those at the pristine air–water interface. *Chemical Science* **10**, 2566-2577 (2019).

19. Colussi AJ, Enami S. Comment on "The chemical reactions in electrosprays of water do not always correspond to those at the pristine air–water interface" by A. Gallo Jr, A. S. F. Farinha, M. Dinis, A.-H. Emwas, A. Santana, R. J. Nielsen, W. A. Goddard III and H. Mishra, Chem. Sci., 2019, 10, 2566. *Chemical Science*, (2019).

20. Gallo A*, et al.* Reply to the 'Comment on "The chemical reactions in electrosprays of water do not always correspond to those at the pristine air–water interface"' by A. J. Colussi and S. Enami, Chem. Sci., 2019, 10, DOI: 10.1039/c9sc00991d. *Chemical Science*, (2019).

21. Nauruzbayeva J, Sun Z, Gallo A, Ibrahim M, Santamarina JC, Mishra H. Electrification at water–hydrophobe interfaces. *Nature Communications* **11**, 5285 (2020).

22. Uematsu Y, Bonthuis DJ, Netz RR. Charged Surface-Active Impurities at Nanomolar Concentration Induce Jones-Ray Effect. *Journal of Physical Chemistry Letters* **9**, 189-193 (2018).

23. Byrnes SJ, Geissler PL, Shen YR. Ambiguities in surface nonlinear spectroscopy calculations. *Chemical Physics Letters* **516**, 115-124 (2011).

24. Agmon N*, et al.* Protons and Hydroxide Ions in Aqueous Systems. *Chemical Reviews* **116**, 7642-7672 (2016).

25. Ruiz-Lopez MF, Francisco JS, Martins-Costa MTC, Anglada JM. Molecular reactions at aqueous interfaces. *Nature Reviews Chemistry* **4**, 459-475 (2020).

26. Jacobs MI, Davis RD, Rapf RJ, Wilson KR. Studying Chemistry in Micro-compartments by Separating Droplet Generation from Ionization. *Journal of The American Society for Mass Spectrometry* **30**, 339-343 (2019).

27. Rovelli G, Jacobs MI, Willis MD, Rapf RJ, Prophet AM, Wilson KR. A critical analysis of electrospray techniques for the determination of accelerated rates and mechanisms of chemical reactions in droplets. *Chemical Science* **11**, 13026-13043 (2020).

28. Pullanchery S, Kulik S, Rehl B, Hassanali A, Roke S. Charge transfer across C–H···O hydrogen bonds stabilizes oil droplets in water. *Science* **374**, 1366-1370 (2021).

29. Wei H, *et al.* Aerosol microdroplets exhibit a stable pH gradient. *Proceedings of the National Academy of Sciences* **115**, 7272-7277 (2018).

30. Colussi AJ. Can the pH at the air/water interface be different from the pH of bulk water? *Proceedings of the National Academy of Sciences of the United States of America* **115**, E7887 (2018).

31. Li M, *et al.* Spatial homogeneity of pH in aerosol microdroplets. *Chem* **9**, 1036-1046 (2023).

32. Roger K, Cabane B. Why Are Hydrophobic/Water Interfaces Negatively Charged? *Angewandte Chemie International Edition* **51**, 5625-5628 (2012).

33. Roger K, Cabane B. Uncontaminated Hydrophobic/Water Interfaces Are Uncharged: A Reply. *Angewandte Chemie International Edition* **51**, 12943-12945 (2012).

34. Jena KC, Scheu R, Roke S. Surface Impurities Are Not Responsible For the Charge on the Oil/Water Interface: A Comment. *Angewandte Chemie International Edition* **51**, 12938-12940 (2012).

35. Beattie JK, Gray-Weale A. Oil/Water Interface Charged by Hydroxide Ions and Deprotonated Fatty Acids: A Comment. *Angewandte Chemie International Edition* **51**, 12941-12942 (2012).

36. Ben-Amotz D. Electric buzz in a glass of pure water. *Science* **376**, 800-801 (2022).

37. Poli E, Jong KH, Hassanali A. Charge transfer as a ubiquitous mechanism in determining the negative charge at hydrophobic interfaces. *Nature Communications* **11**, 901 (2020).

38. Enami S, Mishra H, Hoffmann MR, Colussi AJ. Protonation and Oligomerization of Gaseous Isoprene on Mildly Acidic Surfaces: Implications for Atmospheric Chemistry. *J Phys Chem A* **116**, 6027-6032 (2012).

39. Platt JR. Strong Inference - Certain Systematic Methods of Scientific Thinking May Produce Much More Rapid Progress Than Others. *Science* **146**, 7 (1964).

40.     Otter JA, Yezli S, Barbut F, Perl TM. An overview of automated room disinfection systems: When to use them and how to choose them. *Decontamination in Hospitals and Healthcare*, 323-369 (2020).

41.     Li K*, et al.* Spontaneous dark formation of OH radicals at the interface of aqueous atmospheric droplets. *Proceedings of the National Academy of Sciences* **120**, e2220228120 (2023).

42.     Kakeshpour T, Metaferia B, Zare RN, Bax A. Quantitative detection of hydrogen peroxide in rain, air, exhaled breath, and biological fluids by NMR spectroscopy. *Proceedings of the National Academy of Sciences* **119**, e2121542119 (2022).

43.     Lin S, Cao LNY, Tang Z, Wang ZL. Size-dependent charge transfer between water microdroplets. *Proceedings of the National Academy of Sciences* **120**, e2307977120 (2023).

44.     Xiong H, Lee JK, Zare RN, Min W. Strong Electric Field Observed at the Interface of Aqueous Microdroplets. *The Journal of Physical Chemistry Letters* **11**, 7423-7428 (2020).

45.     Hao H, Leven I, Head-Gordon T. Can electric fields drive chemistry for an aqueous microdroplet? *Nature communications* **13**, 280 (2022).

46.     Heindel JP, Hao H, LaCour RA, Head-Gordon T. Spontaneous formation of hydrogen peroxide in water microdroplets. *The Journal of Physical Chemistry Letters* **13**, 10035-10041 (2022).

47.     Colussi AJ. Mechanism of Hydrogen Peroxide Formation on Sprayed Water Microdroplets. *Journal of the American Chemical Society* **145**, 16315-16317 (2023).

48.     Musskopf NH, Gallo A, Zhang P, Petry J, Mishra H. The Air–Water Interface of Water Microdroplets Formed by Ultrasonication or Condensation Does Not Produce H2O2. *The Journal of Physical Chemistry Letters* **12**, 11422-11429 (2021).

49.     Gallo Jr A*, et al.* On the formation of hydrogen peroxide in water microdroplets. *Chemical Science* **13**, 2574-2583 (2022).

50.     Nguyen D, Nguyen SC. Revisiting the Effect of the Air–Water Interface of Ultrasonically Atomized Water Microdroplets on H2O2 Formation. *The Journal of Physical Chemistry B* **126**, 3180-3185 (2022).

51.     Nguyen D, Lyu P, Nguyen SC. Experimental and Thermodynamic Viewpoints on Claims of a Spontaneous H2O2 Formation at the Air–Water Interface. *The Journal of Physical Chemistry B* **127**, 2323-2330 (2023).

52. Martins-Costa MT, Ruiz-López MF. Probing solvation electrostatics at the air–water interface. *Theoretical Chemistry Accounts* **142**, 29 (2023).

53. Peplow M. Claims of water turning into hydrogen peroxide spark debate. In: *C&EN*). American Chemical Society (2022).

54. Cozens T. Study casts doubt on water microdroplets' ability to spontaneously produce hydrogen peroxide. In: *Chemistry World*). Royal Society of Chemistry (2022).

55. Trager R. Water surprise: microdroplets have potential to produce hydrogen peroxide. In: *Chemistry World*). Royal Society of Chemistry (2019).

56. Riesz P, Berdahl D, Christman CL. Free radical generation by ultrasound in aqueous and nonaqueous solutions. *Environmental health perspectives* **64**, 233-252 (1985).

57. Fang X, Mark G, von Sonntag C. OH radical formation by ultrasound in aqueous solutions Part I: the chemistry underlying the terephthalate dosimeter. *Ultrasonics Sonochemistry* **3**, 57-63 (1996).

58. Hoffmann M, Hua I, Hoechemer R. Application of Ultrasonic Irradiation for the Degradation of Chemical Contaminants in Water. *Ultrasonics Sonochemistry* **3**, 168-172 (1996).

59. Bang JH, Suslick KS. Applications of Ultrasound to the Synthesis of Nanostructured Materials. *Advanced Materials* **22**, 1039-1059 (2010).

60. Mehrgardi MA, Mofidfar M, Zare RN. Sprayed Water Microdroplets Are Able to Generate Hydrogen Peroxide Spontaneously. *Journal of the American Chemical Society* **144**, 7606-7609 (2022).

61. Kakeshpour T, Bax A. NMR characterization of H2O2 hydrogen exchange. *Journal of Magnetic Resonance* **333**, 107092 (2021).

62. Hayyan M, Hashim MA, AlNashef IM. Superoxide Ion: Generation and Chemical Implications. *Chemical Reviews* **116**, 3029-3085 (2016).

63. Bard AJ, Faulkner LR. *Electrochemical Methods: Fundamentals and Applications*, Second edn. John Wiley & Sons (2001).

64. Ivanovic-Burmazovic I. Catalytic dismutation vs. reversible binding of superoxide. *Adv Inorg Chem* **60**, 59-100 (2008).

65. Arunachalam S*, et al.* Rendering SiO2/Si Surfaces Omniphobic by Carving Gas-Entrapping Microtextures Comprising Reentrant and Doubly Reentrant Cavities or Pillars. *JoVE*, e60403 (2020).

# Busting the Myth of Spontaneous Formation of $H_2O_2$ at the Air–Water Interface: Contributions of the Liquid–Solid Interface and Dissolved Oxygen Exposed

Muzzamil Ahmad Eatoo & Himanshu Mishra*

Environmental Science and Engineering (EnSE) Program,

Biological and Environmental Science and Engineering (BESE) Division,

King Abdullah University of Science and Technology (KAUST),

Thuwal, 23955-6900, Kingdom of Saudi Arabia,

Water Desalination and Reuse Center (WDRC),

King Abdullah University of Science and Technology (KAUST), Thuwal, 23955-6900,
Kingdom of Saudi Arabia

Center for Desert Agriculture (CDA),

King Abdullah University of Science and Technology (KAUST), Thuwal, 23955-6900,
Kingdom of Saudi Arabia

*Himanshu.Mishra@kaust.edu.sa

**Section S1: Experimental setup**

**Water microdroplet generation via Sprays**

We adapted the experimental set-up built by Gallo & co-workers[1] to produce water microdroplets. In a co-axial system, water was injected through the inner tube of 100 µm diameter using a syringe pump (PHD Ultra, Harvard Apparatus), and dry $N_2(g)$ was pushed through the outer tube of 430 µm diameter. HPLC-grade water was used and a glass cell (equipped with a tiny opening to prevent pressure build up) was used to collect the microdroplets while minimizing ambient contamination. The water flow rate was set at 25 µL/min, while the gas ($N_2$ or $O_2$) pressure was 100 psi.

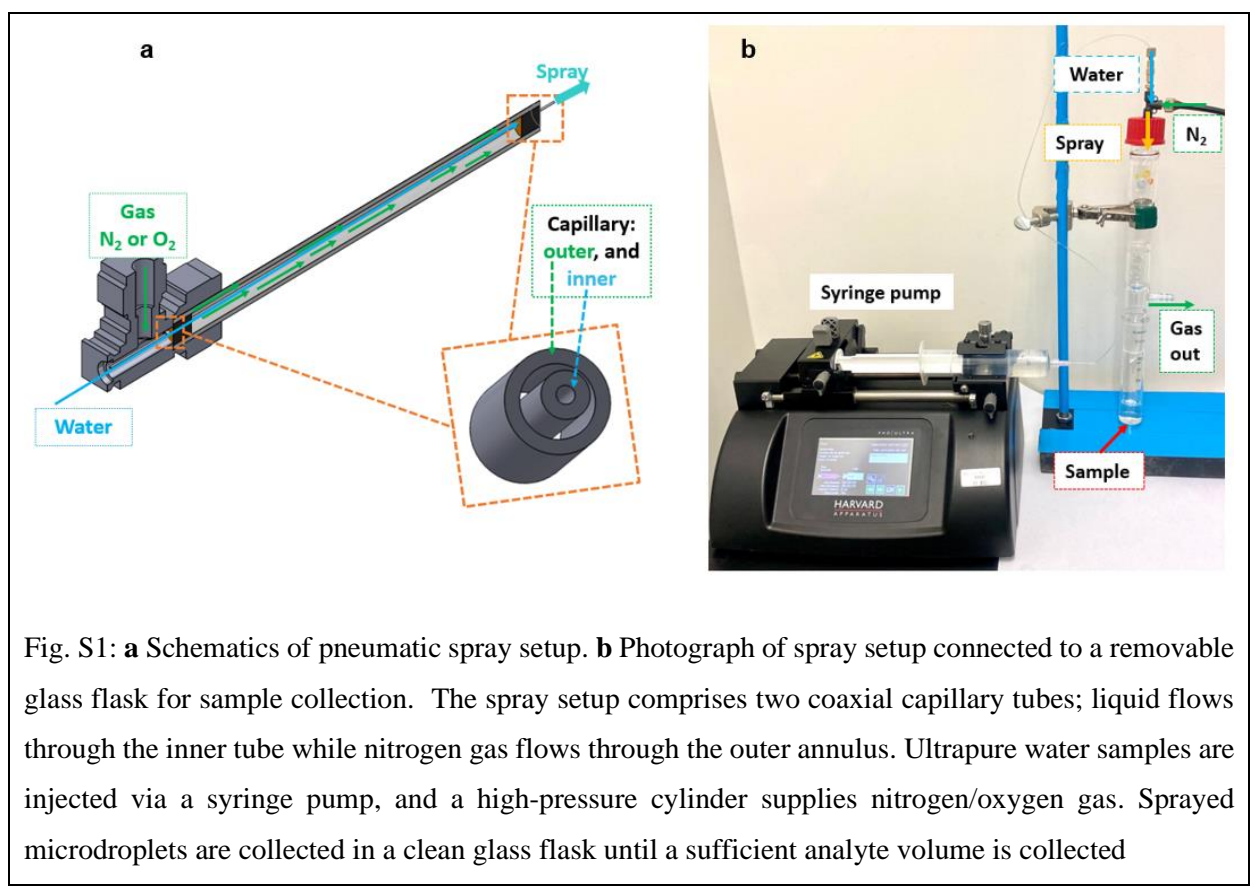

Fig. S1: **a** Schematics of pneumatic spray setup. **b** Photograph of spray setup connected to a removable glass flask for sample collection. The spray setup comprises two coaxial capillary tubes; liquid flows through the inner tube while nitrogen gas flows through the outer annulus. Ultrapure water samples are injected via a syringe pump, and a high-pressure cylinder supplies nitrogen/oxygen gas. Sprayed microdroplets are collected in a clean glass flask until a sufficient analyte volume is collected

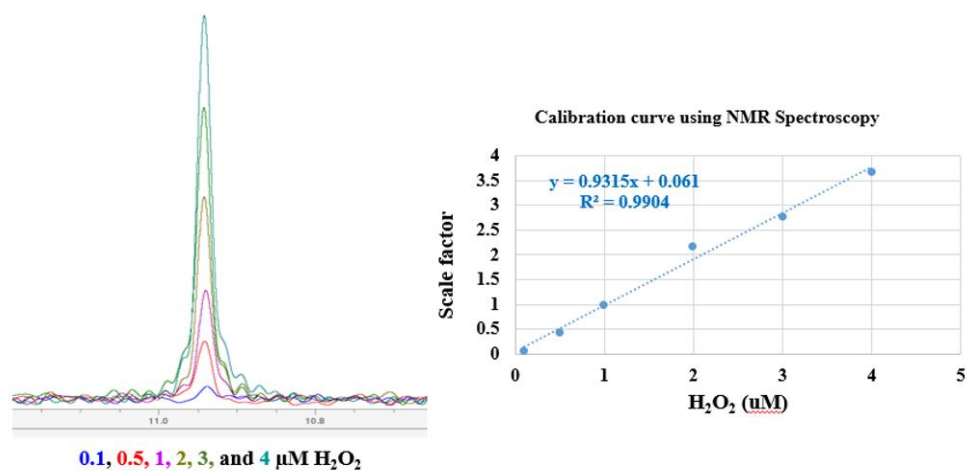

0.1, 0.5, 1, 2, 3, and 4 μM $H_2O_2$

Fig. S2: Quantification of $H_2O_2$ by $^1$H-NMR spectroscopy. Standard samples of known $H_2O_2$(aq) concentrations were prepared by diluting a concentrated 30% (v/v) stock solution using HPLC-grade DI water. All the $^1$H-NMR experiments were conducted at 2 °C. TopSpin 4.2.0 software was used to process the data.

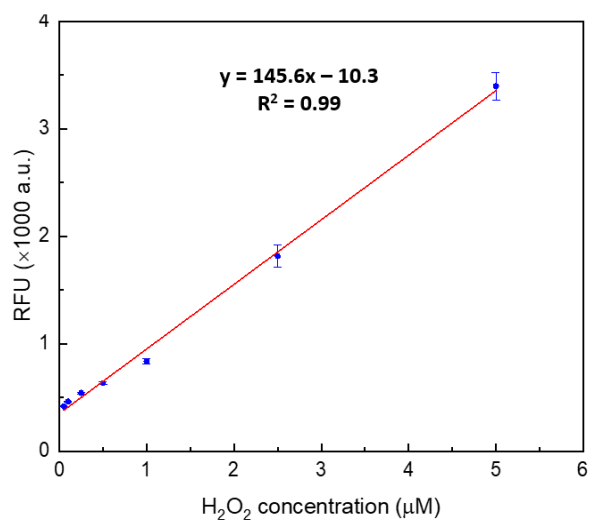

Fig. S3: Representative calibration curve for $H_2O_2$ measurement using hydrogen peroxide assay kit (HPAK)

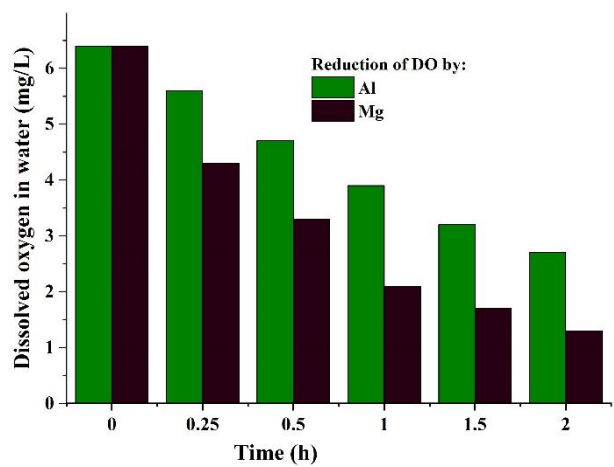

Fig. S4. Reduction of dissolved oxygen concentration in water during $H_2O_2$ formation at the water–solid interface. In a typical experiment, an Al or a Mg pellet was immersed in 5 ml water in a closed vial. The dissolved oxygen concentration was measured over time using a WTW Multi 3320 device with a detection limit of 0.01 mg/l. The results revealed that dissolved oxygen concentration reduced with time as $H_2O_2$ formed. Also, the reduction was faster with a Mg pallet than that with an Al pellet in accordance to their ability to produce $H_2O_2(aq)$.

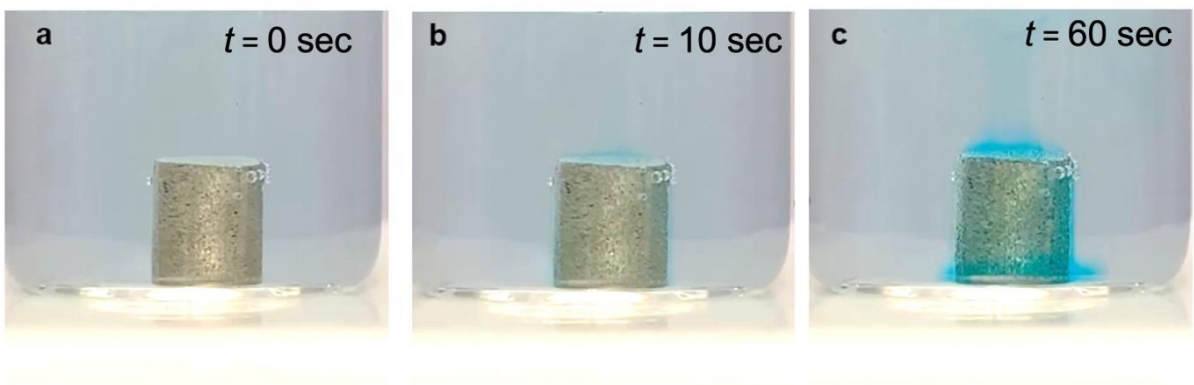

Fig. S5. As Mg pellet is immersed in a 1:1 mixture of water and HPAK, it produces $H_2O_2$(aq) evidenced by the blue fluorescence over time. This proves that this chemical transformation is not depended on the air–water interface, or 'micro' scale water droplets, or ultrahigh (instantaneous) electric fields.

**Table S1**: $H_2O_2$ formation in bulk water as characterized by NMR spectroscopy

| | Solid–Water System<br>Note: water is obtained from a Milli-Q system | Measured $H_2O_2$(aq) concentration after 1 minute<br>(NMR Spectroscopy) |
|---|---|---|
| 1 | 1 mL sessile water droplet placed on a freshly polished aluminum plate (See Supplementary Movie S1) | $1.4 \pm 0.5$ μM |
| 2 | 1 mL water film sandwiched between two freshly polished aluminum plates of size 200× 200 mm$^2$ | $39 \pm 6$ μM |
| 3 | A cylindrical Mg pellet of diameter 1 cm and height 1.5 cm (6.2 cm$^2$) fully immersed in 5 mL of water | $2.5 \pm 0.6$ μM |

**References:**

1. A. Gallo Jr, N. H. Musskopf, X. Liu, Z. Yang, J. Petry, P. Zhang, S. Thoroddsen, H. Im and H. Mishra, On the formation of hydrogen peroxide in water microdroplets, *Chemical Science*, 2022, **13**, 2574-2583.